# Architectural Framework for Large-Scale Multicast in Mobile Ad Hoc Networks

Ahmed Helmy
Department of Electrical Engineering
University of Southern California
helmy@usc.edu

Emerging ad hoc networks are infrastructure-less networks consisting of wireless devices with various *power* constraints, capabilities and *mobility* characteristics. An essential capability in future ad hoc networks is the ability to provide scalable *multicast services*. This paper presents a *novel adaptive architecture* to support multicast services in large-scale wide-area ad hoc networks. Existing works on multicast in ad hoc networks address only small size networks. Our main design goals are *scalability, robustness* and *efficiency*. We propose a *self-configuring* hierarchy extending zone-based routing with the notion of *contacts* based on the *small world graphs* phenomenon and new metrics of stability and mobility. We introduce a new *geographic-based multicast address allocation* scheme coupled with *adaptive anycast* based on *group popularity*. Our scheme is the first of its kind and promises efficient and robust operation in the common case. Also, based on the new concept of *rendezvous regions*, we provide a *bootstrap* mechanism for the multicast service; a challenge generally ignored in previous work.

## 1. Introduction

*Ad hoc networks* are emerging as very interesting architectures to support pervasive mobile wireless devices and are expected to have a significant impact on service paradigms in various vital fields; such as military, disaster relief, bio-sensing, and location-based mobile services. Such networks consist of heterogeneous wireless devices with various power and mobility characteristics. At the same time, *group communication* represents a very important class of applications in future networks. Multicast is the enabling technology for efficient group communication. Providing a scalable architecture for multicast services in large-scale networks has proven to be a challenge for years in the networking community[34][3]. The research challenges are even greater for ad hoc networks, mainly due to lack of infrastructure and the *highly dynamic* nature of wireless nodes and their unexpected *mobility*. The developed multicast service should allow group participants to join and leave at will, and should impose no restrictions on node mobility. It should also provide automatic multicast address allocation and session advertisement. Most existing work on ad hoc networks ignores such requirement.

In this paper, we design a new scalable architecture for multicast services support in large-scale ad hoc networks, the *first* of its kind. Since ad hoc networks are *infrastructure-less*, the developed protocols must be *self-configuring*. Scalability and robustness should be addressed carefully with growth in the size of the network and the number of group participants. Existing approaches for multicast, usually employ flooding or cores for discovery of group participants. These approaches, however, only apply to small networks due to cost of flooding and maintenance of the core tree. Questions are often asked about which routing mechanisms to use, pro-active or re-active routing? Hierarchical or flat? It is usually the case, however, that each of these mechanisms has its strengths and weaknesses depending on network conditions. We attempt to leverage strengths of the different approaches by introducing a *hybrid* approach that *adapts* to network dynamics as a function of *mobility* and *power*.

Our proposed architectural framework utilizes *highly adaptive* mechanisms in the various components of the architecture. Specifically, we design an adaptive zone-based hierarchy augmented by the notion of *contact* nodes to increase network coverage. In addition, we provide novel schemes for multicast service support, based on *adaptive anycast* for resource discovery. We also introduce geographic multicast address allocation to map groups into *rendezvous regions*.

Unlike existing work on multicast in ad hoc networks that mainly addresses local areas (with tens or hundreds of nodes), our work targets *large-scale wide-area* ad hoc networks (with tens of thousands of nodes). Issues of *scalability, service model, robustness* and *efficiency* affect the essence of our architectural design and guide our choice of multicast model. In order to address these issues we discuss the design requirements and present an overview of the architectural components. Then we provide specific mechanisms to support our design goals.

### 1.1 Design Requirements

The main factors driving our design are scalability, the multicast service support, and robustness.

(a) **Scalability**: Unlike most related work that considers tens to hundreds of mobile nodes, our architecture should be able to support large number (tens of thousands) of nodes. We believe that mobile nodes will be pervasive, replacing PCs and cellular phones, with tens of new classes of application supported by mobile wireless devices (e.g., navigation, location-based services). Flat architectures are known not to scale well[7], mainly due to the far-reaching effects of network dynamics; mobility, failures and topological changes. Such effects consume network resources (i.e., bandwidth and power), and lead to recovery delays and increased route oscillations. Hierarchical architectures, on the other hand, alleviate the above problems, as they tend to localize and dampen network dynamics, and scale routing tables using aggregation. Many existing hierarchical architectures are based on *clustering* mechanisms, in which a single node per cluster (called master, cluster-head or parent) is chosen to manage or organize the cluster. Such architectures suffer from single point of failures, in which the failure (or movement) of the master may have severe negative effects on the hierarchy. Furthermore, establishment and maintenance of clusters requires mechanisms for electing the master and mechanisms for joining/leaving clusters, which usually incur a lot of overhead and complexity. We design an architecture that leverages hierarchical advantages while alleviating effects of master and hierarchy maintenance. We adopt a two-level distributed hierarchical architecture. For the first level of the hierarchy we adopt a *zone-based* approach (a variant of the zone routing protocol ZRP[17]), in which each node has its own view of a zone. For the second level, we introduce a novel concept of *contacts* (based on the concept of small-world[18][43]) to enhance a node's view, and aid in route and resource discovery. Also affecting scalability, is the choice of routing protocol. In general, ad hoc routing protocols are either pro-active (i.e., table driven) or re-active (on-demand). Pro-active protocols[6] exchange periodic messages to keep routes up-to-date. Pro-active route discovery has low delay, with significant overhead of periodic route exchange (many of which may become invalid due to mobility). Re-active protocols[10], by contrast, maintain routes on-demand. Re-active route discovery does not incur periodic overhead, but incurs more route discovery delay, which usually involves request broadcasts throughout the network. We believe that neither protocol perform well in all network conditions. We attempt to combine the strengths of both protocols using a hybrid approach. Inside a zone pro-active routing is used to discover routes to nearby nodes, while re-active routing is used for discovery of routes faraway. In addition, *contacts* extend the notion of zone to enhance route discovery.

(b) **Multicast Service Support**: The multicast service model defines conditions for joining/leaving groups and specifies the interaction between the end-nodes (i.e., participants) with the rest of the network. Multicast participants should be able to join, leave or send packets to groups at will. We adopt a model in which participants are not known a priori and are allowed to move freely during a multicast session. In such a model the main problems include *rendezvous*[1] of participants, *service bootstrap* and *multicast address allocation*. We design a novel adaptive anycast architecture for resource discovery to facilitate the *rendezvous* of group participants. Instead of the traditional rendezvous approaches of broadcast and prune[25][5][31] or rendezvous cores[2][4], we introduce a new multicast paradigm based on '*sender push, server cache, receiver pull*' approach, that better fits large-scale ad hoc networks. We also design mechanisms for providing participants with active session information as part of our *bootstrap* architecture, along with a new multicast address allocation scheme based on geographic address allocation. Our architecture requires minimum configuration of nodes.

---

[1] Rendezvous refers to the problem of senders and receivers meeting by knowing information to build the distribution tree.

In our scheme, nodes only need to know a *well-known* session announcement group address and an algorithmic mapping function to map groups into *rendezvous regions (RRs)*. We do not assume existence of unicast routing. Multi-sender groups and geographic scoping are also supported.

(c) **Robustness**: In a highly dynamic environment, such as ad hoc networks, where mobility and crashes are likely to occur, robustness is of prime concern. Being able to adapt to network dynamics to achieve correct behavior and reasonable performance plays a major role in our design. We incorporate mobility and stability models into our hierarchy formation to achieve adaptivity. In addition, our distributed adaptive resource discovery architecture avoids single point of failure scenarios and promises continued operation and graceful recovery during network partitions. We also incorporate path redundancy mechanisms in our multicast routing protocol. Multicast *trees* do not provide sufficient robustness against mobility and failures. We use *mesh* structures for robust delivery. Unlike existing proposals for mesh construction, however, our mechanisms will be designed to build meshes in anticipation of movement. Not only does that achieve better performance, but also provides path redundancy that may be used in case of failures. In addition, mesh branches are activated on-demand to reduce overhead without affecting robustness.

In addition, our mechanisms should be *energy-efficient* and *self-configuring*. Energy-efficiency is one side-effect of scalability. In addition, we attempt to limit communication (one main source of energy consumption) by using localized mechanisms for advertisement and query. Furthermore, in resource-discovery and anycast (where only a single resource is sought), localized broadcast techniques are adopted to reduce overhead. Self-configurability is an inherent feature of our hierarchy formation and resource discovery mechanisms.

### 1.2. Brief Architectural Overview

In order to address the above challenging requirements, we provide an architectural framework based on the following components: (i) Self-configuring adaptive hierarchy formation and adaptation. (ii) The multicast service architecture consisting of: (a) the multicast model, (b) multicast routing, (c) adaptive resource discovery and (d) multicast address allocation.

The rest of this paper is outlined as follows. Section 2 introduces the hierarchy formation and adaptation mechanisms. Section 3 proposes our new multicast services architecture. Section 4 discusses related work. Section 5 presents future work and the conclusions.

### 2. Hierarchy Formation and Adaptation

We provide mechanisms for self-configuring hierarchy formation, based on zone-based routing, augmented by *contacts*, along with description of contact selection mechanisms. Then we propose hierarchy adaptation mechanisms based on link availability and mobility estimation models.

## 2.1 Hierarchy Formation

As was discussed earlier, flat routing architectures do not scale well for wide-area networks, especially for highly dynamic ad hoc networks. Also, hierarchical approaches based on concept of *master* (or *cluster-head*), through which traffic from the cluster funnels, are undesirable.

One possible approach to consider is to use the master for infrequent coordination but not for forwarding packets. One such approach is the Landmark hierarchy (LMH)[21]. Although designed mainly for wired networks, LMH has the ability to self-configure dynamically, without relying on administrative domains and exhibits path lengths and routing table sizes comparable to conventional cluster-based hierarchies (e.g., Internet). The traffic from/to a cluster need not go through the landmark, which adds robustness. LMH, as presented in[21], however, was designed mainly for wired networks, without accounting for mobility dynamics or power constraints. After examination we identified several drawbacks of LMH. LMH employs complex promotion, demotion, and adoption operations for hierarchy maintenance. Furthermore, effects of mobility on the hierarchy were found to be drastic, sometimes leading to total re-configuration of the hierarchy. For example, movement of a high level landmark triggers re-election in its old region, then demotion and adoption for the moving node are triggered (potentially for as many times as the levels of the hierarchy). This may trigger a chain reaction that consumes many resources unnecessarily. In addition, sub-optimal paths may be common due to hierarchical routing. A variant of LMH (called LANMAR) was presented in[27]. LANMAR, however, was based on the premise that group mobility will be dominant in ad hoc networks. Hence, it does not provide mechanisms to solve LMH drawbacks in the general case.

Another approach that avoids complex coordination for architectural setup is the zone routing protocol (ZRP)[17] which defines a zone for every node as the number of nodes reachable within a radius of ***R*** hops away, shown in Figure 1 (a). Inside the zone proactive (intra-zone) routing is used, so nodes obtain routes to all nodes within their zone. To discover nodes outside of the zone reactive (inter-zone) routing is performed by flooding through periphery or *border* nodes of each zone (known as *bordercasting*). ZRP routing overhead depends heavily on the choice of the *zone radius*. If the radius is too small the routing overhead is dominated by reactive overhead, and vice versa. Optimizing such a parameter is nontrivial. A hybrid min-search and traffic adaptive approach is used in[40] but requires relative stability of the network to approach optimality.

ZRP seems appealing, but experiences excessive delays and overheads in large-scale networks, where much of the traffic maybe destined out-of-zone. Therefore, we develop a novel approach that goes beyond the *zone* while maintaining similar simplicity and stability. Our approach is based on a concept we call *contacts*. Contacts, of a certain node *x*, are nodes that previously existed in *x*'s zone but are drifting out-of-zone, and hence have a network view beyond that of *x* or any of its border nodes[2]. While drifting away gradually, *x* may maintain route to (some of) these drifting nodes using low overhead (since these drifting nodes were in *x*'s zone and are close to its border nodes). Figure 1 shows a simple illustrative example of zoning, contacts and effects of mobility.

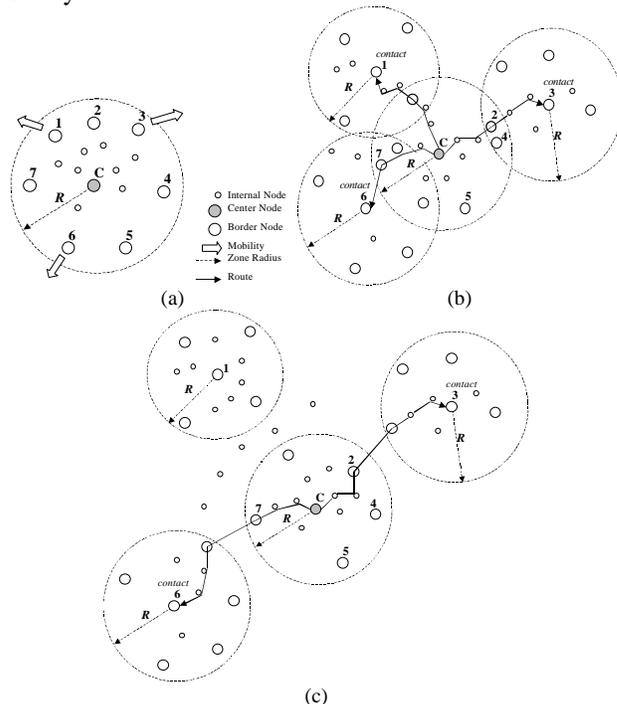

Figure 1. Example of zoning, contacts and effect of mobility: (a) Zone for center node *C* is shown (with radius *R*). Border nodes are numbered (1-7). Nodes 1,3 and 6 are moving/drifting out of zone. (b) Radii for the drifting nodes are shown. *C* stays in *contact* with the drifting nodes, which enables it to obtain better network coverage with low overhead. (c) After moving away, contact nodes drift up to a point where their zones no longer intersect with *C*'s zone. In this example, *C* maintains contact with those nodes not more than (2*R*+1) hops away, i.e. nodes 3 and 6, and loses contact with node 1 as it drifts farther than the contact zone.

The concept of contacts relates to the concept of *small world* graphs, where Watts[18] observes that small world graphs can have low average path length with high clustering. In other words, introducing a small number of far away links (i.e., short cuts) leads to significant reduction in path length. He also found that picking random far away nodes may lead to small world graphs, depending on the probability of choosing the far link. We believe, however, that picking contacts at random is undesirable, since this may lead to unpredictable overhead for contact route discovery and maintenance. In addition, knowing mobility characteristics and stability of a node helps identify better (more useful) contacts. We *take advantage of node mobility* and pick contacts from those nodes drifting away from the zone. The eventual characteristics of the formed graphs are function of time and depend on the initial choice of contacts and the node mobility.

---
[2] It seems this is one of the few concepts that actually takes advantage of mobility. Mobility is often viewed as a disadvantage, and for good reasons, but we think it should also be taken advantage of, when possible.

*Contacts* play an important role in route and resource discovery in our architecture, as will be explained later. A node should choose its contacts carefully to attempt to maintain a relatively long and useful contact as long as the contact route is kept. The *contact list*, maintained by a node, changes *adaptively* as the network conditions change. We further discuss how the contact list is chosen in the next section.

**2.2 Hierarchy Adaptation**

The architecture presented thus far provides a framework for hierarchy formation. Such hierarchy should be highly adaptive to network dynamics and conditions by integrating concepts of *mobility* and *energy*. To achieve this, we use mechanisms that integrate measures of stability and power. We use such mechanisms for hierarchy formation and route selection. In hierarchy formation, the zone radius and contact selection may be adapted dynamically to establish certain stability measures for zones and contacts. Furthermore, each node should adapt its behavior to achieve desirable collective performance. For example, number of contacts chosen by a node should be a function of the contacts already established by other nodes in its zone. Increase in number of contacts increases bandwidth and must be done only as necessary.

We devise a mechanism for contact selection in which a node chooses its contacts with probability $p$, as a function of the border's mobility, the number of zone contacts, energy and activity. One simple model is to have $p$ proportional to the energy estimates $E_{est}$ of the node and the contact, their relative stability $S_{est}$, and the activity level of the node $A_{est}$ measured as rate of discovery requests. Also, $p$ is inversely proportional to the number of zone contacts $Z_{est}$. Hence, $p \propto \frac{E_{est} S_{est} A_{est}}{Z_{est}}$. These quantities are locally measured and may be piggybacked on intra-zone pro-active messages. $E_{est}$ includes energy estimates at the node choosing the contact, and the contact node drifting out of zone. To accommodate heterogeneous nodes, the estimates should include energy left $E_{left}$, and the drainage $\Delta E$. Thus, a simple equation for the energy estimate is: $E_{est} = \left(\frac{E_{left}}{\Delta E}\right)_{node} \left(\frac{E_{left}}{\Delta E}\right)_{contact}$.

Stability estimates may be derived from adaptive availability and mobility models. In[7] a scheme was proposed to measure link and path *availability*. The model used is $(a,t)$, where $a$ is the probability that a link will be available for time $t$, and was utilized to build adaptive clustering schemes for ad hoc networks. The basic idea is to build stable clusters in which internal route availability is probabilistically bounded. The proposed model is a random walk-based model that determines the conditional probability that the nodes will be within range of each other at time $t_0+t$ given that they are located within range at time $t_0$. Another mobility metric may be derived from signal strengths, using a propagation model; $\frac{RxPwr}{TxPwr} \propto \frac{1}{d^n}$ where $RxPwr$ ($TxPwr$) is received (transmitted) signal power, and $d$ is the distance between transmitter and receiver. By measuring the signal strength between two consecutive packets received from the same transmitter a *relative mobility* metric is defined as $\log \frac{RxPwr_{new}}{RxPwr_{old}}$, if negative high value then nodes are moving away quickly, and vice versa. This metric was used in[45] to aid in choosing cluster-heads.

The above adaptive mechanisms are also used in the multicast service model and resource discovery in section 3. In such case, contact's choice should also be affected by the capabilities of the contact node (e.g., global positioning system (GPS) capability, or being a session or sender discovery server).

**3. Multicast Service Architecture**

Providing a scalable multicast service architecture is the focal point of our research. The hierarchical architecture proposed thus far provides the basis for efficient support of our multicast service model, presented in this section. As the mechanisms for our multicast model unfold, the essential role played by the adaptive hierarchical architecture will become very clear in support of multicast routing and resource discovery. In general, existing work on ad hoc multicast concentrates on multicast routing in small to medium size networks. Earlier work particularly focused on mechanisms to establish distribution trees (or meshes) between senders and receivers in small-scale networks, mainly using periodic broadcasts (either from the sender[25] or receiver[28] side) or relying on *cores*[15]. In this paper, however, we address multicast in large-scale ad hoc networks, and propose new schemes for *bootstrapping* multicast services, by providing efficient *resource discovery* using *popularity-based adaptive anycast* and *geographic multicast address allocation*. We are not aware of existing or on-going work on multicast for wide-area ad hoc networks. Furthermore, we do not know of other work for *bootstrapping* multicast service, resource discovery or multicast address allocation in ad hoc networks.

Our multicast service architecture consists of three main components; (a) the multicast model (i.e., how senders and receivers meet), (b) multicast routing (i.e., establishing multicast distribution paths), and (c) *adaptive resource discovery* architecture using anycast and geographic multicast address allocation.

**3.1. The Multicast Model**

The basic premise for scalable multicast is that sources do not know who/where receivers are a priori. This model enables any node to join or leave a multicast group at any point in time. Hence, one of the main components of multicast is a mechanism for group participants to *meet* or *rendezvous*. Traditionally, this problem has been addressed, in wired and ad hoc networks, using one of two approaches; *broadcast-and-prune*, or *rendezvous cores*. In the former[5][31][25], a participant (usually the sender) announces its presence by broadcasting data packets (or control messages) throughout the network. Network nodes not interested in the group send prune messages to stop the flow of packets (or simply do not respond in case of control messages broadcast). These broadcasts are periodic to capture network and membership dynamics. It has been shown that such model is best suited for small to medium size networks with densely populated groups, but does not scale for wide-area networks[48]. The rendezvous cores approach, by contrast, uses *explicit join* mechanisms to avoid periodic broadcasts. Participants join (or send packets) towards a common core, which relays the packets from the senders to the receivers using a shared tree (or mesh)[2][15]. This approach suffers from problems of traffic concentration and single point of failure scenarios for the core. These problems may be alleviated by using multiple cores and dynamic core election mechanisms. We believe, however, that the major research problem associated with the core approach is the *core bootstrap* and consistency problem. How do participants know the core's address/location? Senders and receivers need to maintain a consistent view of the cores in order to *meet*. This problem was addressed for wired networks[4] within a single domain, and uses a flooding to disseminate core-to-group mapping. This scheme does not scale well for wide-area networks and its convergence performance degrades with the size of the network. For ad hoc networks these problems are exacerbated by network dynamics and node mobility. Hence, these approaches are not suitable for ad hoc networks.

We propose a new multicast model. We refer to our model simply as *sender push, server cache, receiver pull* model. Unlike previous work on ad hoc multicast that requires periodic broadcasts throughout the entire network, our scheme incurs less overhead, and only when necessary as necessary and as localized as possible. We introduce the notion of *sender discovery servers (SDS)* to aid in sender location and information dissemination. As shown in Figure 2, a sender sends an Advertisement (*Adv*) using *localized broadcast*. *SDS*s receiving the *Adv* store this information. Receivers send join requests toward the sender based on *backward learning*; every node forwarding the *Adv* adds its address to the message to construct a path back to the source. Other, farther, receivers not receiving the *Adv* message attempt to find a nearby *SDS*, first by checking in their own zone (*SDS*s advertise in their zone), then by checking with their contact list. If *SDS* is found it is queried for group information and responds with a join reply, with approximate source location or possible routes (if available at the *SDS*). Depending on the quality of the provided routes (if any), the querying receiver(s) may opt to use these routes or use zone/contact search for other routes (eventually, geographic routing may also be used for route discovery as in LAR[24], as described later). If *SDS* is not found, a receiver may send a localized broadcast to discover other nearby receivers of the group. If this process fails then the receiver uses a fallback mechanism, described later in this section. Once the information about the group/senders is available, the route discovery/construction is initiated. An illustrative example is shown in Figure 2.

### 3.2 Multicast Routing

Establishing multicast distribution paths for ad hoc networks has been shown to be more robust using *mesh* structures, as opposed to conventional *tree* structures[16]. For single-source groups or when sources are sparsely distributed, however, even a mesh does not provide the desired path redundancy. Local recovery mechanisms[44] may be used to alleviate such a problem. Conventional receiver-initiated multicast schemes setup reverse path forwarding (RPF) trees[2] due to dependence on unicast routing. For ad hoc networks, however, using RPF paths is not desirable due to possible path asymmetry due to wireless channels. We do not use RPF nor do we rely on existence of a unicast routing protocol.

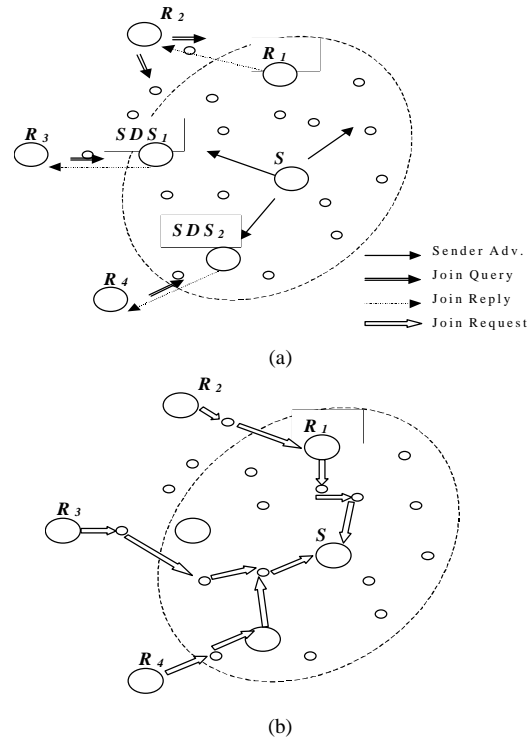

**Figure 2.** Multicast service scenario. (a) Sender $S$ becomes active and broadcasts an advertisement (Adv.) locally (in shaded region). Sender discovery servers $SDS_1$, $SDS_2$, and receiver $R_1$ receive the Adv. When new receivers join the group, they try to find: (i) a sender discovery server, (ii) nearby members of the group. Receiver $R_2$ finds $R_1$ using local broadcast, while $R_3$ and $R_4$ find $SDS_1$ and $SDS_2$. (b) Once sender information is obtained, receivers send Join Requests to build multicast distribution paths.

We propose to use mesh structures with local recovery mechanisms, similar to those used in[25] and [44]. In addition, multiple paths may be selected by the receiver (during the discovery process) to increase robustness of the multicast distribution mesh. Depending on the number of senders in the group and their location (if available), in addition to the mobility of the receiver, a receiver may opt to choose several stable paths to join the group (when propagated, route information includes stability metrics). The receiver sets an 'active' flag in only one join to activate only one path at a time. Only active paths forward data packets. This reduces packet transmission overhead (a very significant factor in ad hoc networks) while maintaining robustness. If performance or stability of the active path degrades, or local recovery fails, the receiver may activate another path with high stability. Also, when the receiver moves, it may activate another path containing one of the new neighbors, thus achieving fast handoff using the concept of multicast-based mobility[1]. Our investigations show that, on average, a moving node traverses 2.5 hops to reach the nearest point of the multicast distribution structure in very large networks (with up to 5000 nodes)[3]. Moreover, mobility prediction[26] may be used to achieve advance joining, further reducing the effects of mobility handoff (i.e., mobility of the receiver in ad hoc networks). Rules for activating/de-activating branches of the distribution structure should be carefully selected to avoid black holes. For example, if *any* member exists downsrtream a branch then the branch must be activated. In order for a branch to be inactive, all downstream branches must be inactive.

So far, we have assumed that the receivers' efforts in searching for group information are successful, using the above localized search techniques. In case of sparse groups, where participants are far apart, or in case of multi-sender groups, where information obtained from other group members may not be complete, a *fallback* mechanism should be used. Such mechanism should be efficient, avoiding frequent global flooding and should be adaptive to membership dynamics. To achieve this, we introduce a novel *bootstrapping* anycast architecture for multicast service in ad hoc networks, discussed next.

### 3.3. Resource Discovery and Multicast Address Allocation

The major research challenge for multicast resource discovery (i.e., discovery of group address, senders), is the lack of any (centralized or distributed) infrastructure to hold and distribute such information. Inter-domain multicast for wired networks[34][3] utilizes the AS hierarchy of the Internet and uses BGP extensions to distribute multicast routes that map group prefixes into root-domains (established by the multicast address allocation). In turn, receivers join towards the root-domain and senders send their packets towards it. Intra-domain multicast is used within the domains and is built on top of unicast routing. All these infrastructures (unicast routing, AS hierarchy) do not exist in ad hoc networks.

An architecture for *anycast* routing in the Internet was recently proposed in[20]. Utilizing hierarchical routing and aggregation. This work provides a scalable mechanism to discover members of anycast groups that are closer to the requester than other members. The architecture identifies two mechanisms, a low overhead mechanism, using default routes, for non-popular anycast groups (in which case the request is routed to the home domain, derived from the anycast address itself), and another mechanism for popular anycast groups that caches routes for nearby members.

The only global infrastructure we can probably utilize in ad hoc networks is *geographic location*. Based on geographic multicast address allocation, we devise a new adaptive anycast architecture as follows. The multicast address space is broken into prefixes. Each multicast address prefix is assigned to a geographic region called the *rendezvous region (RR)*[4]. Nodes located in the *RR* have a collective responsibility of maintaining information about the groups belonging to the group prefix assigned to their current region. Since it is 'collective' and could be done by only a small subset of *SDS* nodes (say 3-7 uncorrelated nodes) we can use a probabilistic promotion scheme for nodes to become SDS for the group prefix. Each node decides locally whether it will become a *SDS* based on its own configuration (some nodes maybe configured as servers), capabilities (e.g., GPS), power and stability estimates. If so, it obtains its (approximate) geographical location[5] and determines the group prefix to which its current location maps, using algorithmic mapping in the general form of $f(x_1 \text{ to } x_2, y_1 \text{ to } y_2) = G_{prefix}$, or similar[6]. At that point, the node acts as a member of the anycast group of *SDS*s responsible for $G_{prefix}$, and advertises this information in its zone and to its contacts[7]. Other *SDS*s for the same prefix reply to update the new *SDS* (the reply is localized to reduce overhead). As *SDS* nodes move out of the RR for the corresponding $G_{prefix}$, they advertise their latest group information and leave message to the RR (using geocast[32], for example), which increases the probability of other nodes promoting themselves to become *SDS*s. This ensures constant replenishing of the pool of *SDS*s serving as members of the anycast group for that *RR*.

The above scheme requires approximate knowledge of geographic location. We do not assume that all nodes are GPS capable. We do assume, however, that nodes are heterogeneous; i.e., *some* nodes are GPS capable, while others use GPS-less techniques[46][47] to discover their approximate relative location.

---

[3] These results were obtained for Internet topologies. We shall investigate these measures in the context of ad hoc networks.

[4] This does not necessarily imply, however, that these groups are geographically limited to that region.
[5] Geographic location update need not be done frequently, only when a node moves noticeable distances. Afterall, this information is approximate and is used for distributed resource discovery (not for forwarding packets).
[6] Nodes express their desire to use multicast to their neighbors, zone or contacts, which will result in a reply with the well-known algorithmic mapping function. Our scheme allows for changing this mapping to another well-known mapping (very infrequently, though).
[7] Alternatively, a node may advertise this information using localized broadcast, or geographically scoped broadcast (where only nodes, within a specific region, broadcast the packets), or Geocast[32].

**Session Initiation** A node initiating a multicast session is expected (without necessity) to use the above algorithmic mapping to obtain a multicast address that maps into a geographical vicinity as its *RR*. In any case, group/session initiation requests/updates are sent to the *RR* to avoid collisions in multicast address allocation. When a new sender of a group belonging to $G_{prefix}$ becomes active it performs a localized broadcast (as described earlier) and, if far from its *RR*, issues an update to *RR*. The requests and updates are sent using *lollipop-LAR* (our modified location aided routing (LAR) [24]) to improve scalability. In *lollipop-LAR*, a far away sender chooses a contact that is closer to the *RR*. If the distance between the contact and the *RR* is less than a limit $l$, the contact sends the request/update to the *RR* directly using LAR, otherwise it chooses one of its contacts closer to the *RR*, and so on. An illustration is shown in Figure 3.

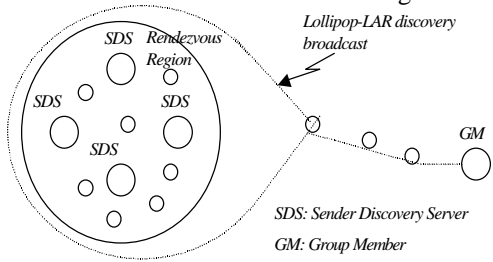

Figure 3. A group member uses algorithmic mapping to obtain the *rendezvous region* of the group, then uses *lollipop-LAR* to reach the region and contact a sender discovery server to obtain the group information.

When a node joins the group, it first attempts localized search for *SDS*. If this fails it sends/geocasts a join query to the *RR* using lollipop-LAR. A join reply is issued by a *SDS* in the *RR* and follows the backward route created by the join query (this path is only used for the reply so it need not be the shortest path). Once the receiver has group/sender information it sends a Join request as described earlier to establish multicast distribution path(s).

**Popularity-based Dynamic Adaptation** If used for the common case, geocasting to the *RR* may incur a lot of overhead to maintain group information. Hence, our multicast service and resource discovery paradigm should adapt to dynamics of membership, to achieve better performance for popular groups. As explained above, joining or sending to groups entail local advertisement (Adv) or query. This gives indications of the popularity of the group in the vicinity of the participants. Nodes receiving Advs and queries, those that are willing to become *SDS* for that group (based on their configuration, stability and capability), estimate the popularity of the group. The initial estimate is based on Advs/queries heard. If this initial estimate exceeds a threshold $pop_{query-th}$ a local group query is sent by the candidate-*SDS* to its own zone (using pro-active routing updates readily sent) and to its contacts. Response to this group query gives a better estimate of group popularity $Grp_{est}$, in addition to information about existing *SDS*s nearby

$SDS_{est}$. Popularity estimate $pop_{est}$ is be obtained as $pop_{est} \alpha \frac{Grp_{est}}{SDS_{est}}$. If $pop_{est} > pop_{th}$ where $pop_{th}$ is the popularity threshold, then a node advertises itself as *SDS* for the group to its zone and contacts, and contacts the *RR SDS* for updates (using lollipop-LAR). Future nearby join queries for the group reach the local *SDS* and are answered locally, reducing overhead and delay. Furthermore, this adaptive mechanism also achieves better robustness and continued operation during network partitions, when RR is unreachable. Illustration of popularity-based adaptation is shown in **Figure 4**

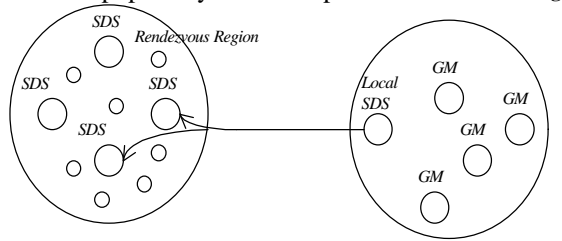

Figure 4. If the group becomes popular in a region away from the rendezvous region a *subgroup* is formed and a local sender discovery server is chosen to contact the rendezvous region.

**Discussion** We note that the probability of success of the localized search is affected mainly by two factors. The first is the group address, obtained during session initiation, which decides the location of the *RR*, and in turn determines the location of the *RR-SDSs*. The second factor is the nearby popularity of the group, which decides the promotion of nearby *SDSs*. Many of the offered services are expected to be *location-based* services, meaning it is targeted to a specific location. Hence, these groups will tend to be popular within certain locations more than others. In addition, initiators are expected to choose group addresses that have *RR* in the geographical vicinity. Both these factors increase the probability of success for localized search, and lead us to believe that, in the common case, our architecture is capable of high performance, with low overhead and low delays. Our schemes work for global as well as locally-scoped groups.

One essential question to ask here is 'how do participants know about newly initiated sessions and their properties?' This can be provided using the same scheme provided above, as follows. When multicast participants express their interest in multicast service, they obtain the algorithmic mapping function (described above) as well as a *well-known session advertisement* group address. Like other groups, this well-known group has its own *RR*. As groups are initiated, they are updated at the *RR-SDS* and the local *SDSs* (if any), then information about new sessions is obtained as above. This provides a *bootstrapping* mechanism essential for providing the multicast service.

## 4. Related Work

Related work lies in the areas of ad hoc routing (unicast and multicast), hierarchy and cluster formation, anycast architectures, inter-domain multicast, and geocasting. In the area of unicast ad-hoc routing, protocols are generally classified as either pro-active (or table-driven) or re-active (or on-demand) protocols. Pro-active protocols include DSDV[6], CGSR[8], and WRP[9], and rely upon routing updates to maintain consistency of route information. Reactive protocols include AODV[10], DSR[11], TORA[12], ABR[13], and SSA[14], and create routes only when required by the source node. One feature of SSA is that it selects routes based on the signal strength between nodes. Fisheye state routing (FSR)[49] is used to reduce routing update overhead of link state routing. For a node, the route update frequency to a certain destination is inversely proportional to the distance (in hops) of the destination. That is, routes to nodes within a small distance are sent to neighbors with higher frequency than routes to far away nodes. This reduces route overhead (not table size) and reduces accuracy of routing with distance. Routing efficiency decreases and delays increase, however, with dynamics of mobility, and routing table size grows linearly with network size[27]. CEDAR[29] uses a simple approximate algorithm for building a core graph consisting of the minimum dominating set of nodes. The mechanism was designed for a network of small (10s of nodes) to medium (100s) size network. A variant of CEDAR[30] was used for multicast by joining to the core graph. The effects of mobility and concentration on the core graph, however, were not clear in the study. Other works on multicast ad-hoc routing in[15][16] generally extend existing multicast routing for the Internet, such as PIM-SM[2]. Other recently proposed multicast ad hoc routing protocols include tree-based and mesh-based protocols. Tree-based protocols include AMRoute[22] and AMRIS[23]. AMRoute creates a bi-directional shared core-based tree using unicast tunnels, it uses virtual mesh links for tree creation and needs unicast, but incurs temporary loops and chooses sub-optimal routes with mobility. AMRIS uses a shared tree and an ID number per node, does not need unicast, broadcasts new-session messages and uses beacons to detect disconnection and re-joins to potential parents. However, it uses the expanding ring search mechanism for branch re-construction due to node failure, which does not scale well. Mesh-based protocols include ODMRP[25] and CAMP[35]. CAMP uses a shared mesh, and all nodes keep membership, routing and packet information. New members use expanding ring search to find other member neighbors. However, CAMP needs a special unicast protocol for its proper operation. ODMRP floods packets within mesh, but follows an on-demand policy for establishment and update of the mesh. It uses request and reply phases, broadcasts source announcements, and does not require unicast routing. The mesh is created when join requests from multiple receivers are sent to multiple-sources. Hence, for sparse groups or single-sender groups ODMRP may not be robust. A local route recovery scheme[44] may be used to address this problem. In our routing protocol, we utilize the concept of mesh construction and local recovery, but we attempt to avoid floods for resource discovery, using a contact-based query approach. Moreover, we allow on-demand-activated multiple paths to be constructed to a single source, to increase robustness and achieve better handoff performance during mobility. Multi-path routing was proposed[37] for parallel data distribution. We will leverage mechanisms provided for multi-path discovery, but we use multiple-paths for multicast differently, by activating on path at a time on-demand.

Recently, a scalable anycast architecture (discussed in Section 3.3) was proposed for the Internet[20], and was discussed earlier. Work on location-based routing was presented in location-aided routing LAR[24], Geocast[32][33]. We use concepts of geocasting for route and resource discovery. We modify it for scalability using lollipop shaped regions with the aid of contacts. We also leverage work on GPS-less positioning[46][47] to determine relative approximate positions of the nodes, assuming GPS capability in *some* nodes.

One of the earliest works on self-configuring hierarchical architectures includes work on the landmark hierarchy. Each node has a level in the hierarchy and a radius *r* associated with that level. Each node advertises information about itself to nodes within *r* hops. So, a node receives advertisements from nearby nodes that are the lowest level of the hierarchy, and faraway nodes that are at higher levels of the hierarchy, and so on. In[38] landmark hierarchy is used to form an object location architecture for sensor networks. Hierarchy levels are self-configuring and may be adapted using a promotion/demotion scheme. Drawbacks of landmark hierarchy were discussed in Section 2.1. LANMAR[26][27] uses the landmark hierarchy concepts to establish hierarchy in ad hoc networks. However, landmarks are used for sets of nodes moving together as a group to reduce routing information exchange. Other hierarchical ad hoc routing include the zone-based hierarchical link state (ZHLS)[39]. ZHLS is a GPS-based routing protocol for ad hoc networks, where a network is divided into non-overlapping zones. A node only knows node connectivity within its zone and the zone connectivity for the network. This architecture does not use cluster head to mitigate traffic concentration, reduce routing protocol control/message exchange overhead and avoid single point of failure. It uses zone ID and node ID for routing. However, the zone map is defined by design for interzone routing, and hence does not adapt to network changes and dynamics. Another protocol called the zone routing protocol (ZRP)[17][40] was discussed in Section 2.1. In[19] the ZRP approach is coupled with geographic (geodesic) routing for remote routing. In[7][41][42] the link availability model is proposed (discussed in Section 2.2). The authors suggest to use it with a cluster based approach, in which a parent is selected based on the availability model to increase the lifetime of the cluster. Parent selection and cluster dynamics may complicate our architecture. Instead, we propose to incorporate the availability model with our modified ZRP approach and to use it for determining zone size, and in choosing contacts.

In the Internet, Hierarchical PIM[34] was proposed is an inter-domain architecture based on the PIM-SM protocol. It suggests a hierarchy of rendezvous points (equivalent to cluster heads) to communicate between multicast domains. The BGMP architecture[3] was proposed for hierarchical inter-domain multicast. It uses a bi-directional shared tree and the notion of a root domain. The problem of multicast address allocation is coupled with BGMP for the choice of the root domain. The same study proposes the MASC scheme for multicast address allocation. Such problem is still active in research.

## 5. Conclusion and Future Work

We have presented the first *architecture for multicast service* support in *large-scale ad hoc* networks. Our architecture is based on the zone-based routing concept, but extends it using our novel concept of *contacts* to increase zone coverage and reduce route and resource discovery overhead. Our mechanisms are *self-configuring* and *highly adaptive* to network dynamics and *mobility*, which renders our architecture more *robust, efficient* and *scalable*. We also provide the first architecture for *adaptive anycast* in ad hoc networks and the first scheme for geographic based multicast address allocation based on our new concept of *rendezvous regions*. We hope that these mechanisms can potentially provide the *resource discovery* component for a wide-array of future applications and middle-ware. Our future work includes thorough evaluation of the architecture and fine-tuning of the mechanistic parameters presented herein.

More specifically, we plan to study the characteristics of the resulting network/small-world graphs due to the use of *contacts* and the probability *p* of choosing a contact. These characteristics, we expect, will be function of time and depend on the initial choice of contacts and the node mobility. Furthermore, our plans include thorough analysis of the performance of the architecture as function of the popularity threshold $pop_{th}$ and other popularity-based parameters. Our hope is that our work provides a framework for further research in the area of multicast (and other areas) in large-scale ad hoc networks.